\documentclass[onecolumn,secnumarabic,amsmath,amssymb,balancelastpage,nofootinbib]{article}

\usepackage[e]{esvect}
\usepackage{color}         % produces boxes or entire pages with colored backgrounds
\usepackage{graphics}      % standard graphics specifications
\usepackage{graphicx}      % alternative graphics specifications
\usepackage{epsf}          % old package handles encapsulated post script issues
\usepackage{bm}            % special 'bold-math' package

\usepackage{natbib}
\usepackage{amssymb}
\usepackage{amsmath, esint}
\usepackage{mathrsfs}
\usepackage{framed}
\usepackage{bigints} % for big integral symbol
\usepackage{enumitem}
\usepackage{pifont}
\usepackage[capbesideposition={left,center},facing=yes,capbesidewidth=8cm,capbesidesep=quad]{floatrow} % for side captioning
\usepackage{setspace} % for double or single spacing

\usepackage[none]{hyphenat} % removes hyphenation

\usepackage[colorlinks=true]{hyperref}  % this package should be added after all others

\definecolor{darkred}{rgb}{0.6,0,0}
\definecolor{darkgreen}{rgb}{0,0.5,0}
\definecolor{darkblue}{rgb}{0,0,0.6}
\hypersetup{ colorlinks,
linkcolor=darkblue,
filecolor=darkgreen,
urlcolor=darkgreen,
citecolor=darkred }

\newcommand{\del}[0]{\ensuremath{\vec{\nabla}}}

\begin{document}

\sloppy % stops it from having words stick out to the right because I forbade hyphenation

\bibliographystyle{authordate1}

%\numberwithin{equation}{section} %sets equation numbers <chapter>.<section>.<index>
%\numberwithin{figure}{section}

\title{\vspace*{-35 pt}\Huge{Putting Positrons into\\Classical Dirac Field Theory}}
\author{Charles T. Sebens\\California Institute of Technology}
\date{December 10, 2019\\ arXiv v.4\\\vspace{10 pt}Forthcoming in \emph{Studies in History and Philosophy of Modern Physics}}

\maketitle
\vspace*{-20 pt}
\begin{abstract}
One way of arriving at a quantum field theory of electrons and positrons is to take a classical theory of the Dirac field and then quantize.  Starting with the standard classical field theory and quantizing in the most straightforward way yields an inadequate quantum field theory.  It is possible to fix this theory by making some modifications (such as redefining the operators for energy and charge).  Here I argue that we ought to make these modifications earlier, revising the classical Dirac field theory that serves as the starting point for quantization (putting positrons into that theory and removing negative energies).  Then, quantization becomes straightforward.  Also, the physics of the Dirac field is made more similar to the physics of the electromagnetic field and we are able to better understand electron spin.
\end{abstract}

\setstretch{1.1}
\tableofcontents
\setstretch{1.3}
\newpage

\section{Introduction}\label{introsec}

In physics, classical Dirac field theory has not received the attention that it deserves.  There are a number of reasons why.  One reason is that, unlike our classical theory of the electromagnetic field, classical Dirac field theory does not provide an accurate description of macroscopic physics.  That is because classical Dirac field theory does not emerge as a classical limit from our quantum theory of the Dirac field.\footnote{See \citet[pg.\ 221]{duncan}.}  Another reason is that---because the Dirac field operators in the quantum field theory anticommute---the classical Dirac field is sometimes treated as Grassmann-valued.  Grassmann numbers are somewhat odd mathematical devices that are generally unfamiliar to physicists before encountering their use in the path integral formulation of quantum field theory for fermionic fields (like the Dirac field).  A third reason that classical Dirac field theory has been neglected is that the equations of the theory were initially given an entirely different interpretation by Dirac and others.  On this alternative interpretation, the equations are viewed as part of a relativistic quantum theory for a single electron.

Classical Dirac field theory is worthy of study because of the foundational role it plays in quantum field theory.  Quantum electrodynamics (the quantum field theory that describes photons, electrons, and positrons) can be arrived at by quantizing the classical electromagnetic and Dirac fields.  If we set aside interactions between these two fields (as we will do throughout this paper), then we can describe this procedure as separately quantizing classical electromagnetism in order to get a quantum theory of the electromagnetic field (a quantum field theory for photons) and quantizing classical Dirac field theory in order to get a quantum theory of the Dirac field (a quantum field theory for electrons and positrons).\footnote{Classical electromagnetism plays two important roles in relation to quantum field theory: it serves as both the classical field theory that gets quantized to arrive at a particular quantum field theory and as a classical theory that approximates this quantum field theory in appropriate circumstances.  Classical Dirac field theory plays the first role but not the second.  You can build a quantum field theory by quantizing the classical Dirac field, but classical Dirac field theory does not accurately approximate this quantum field theory.}

The process of quantizing the Dirac field is not smooth.  One must make various modifications on the way to quantum field theory, such as redefining the operators for energy and charge.  In this paper, I will argue that the reason these modifications are necessary is that we start with a classical theory of electrons (where the energy is not necessarily positive) and are then somehow trying to get out a quantum field theory of electrons and positrons (without negative energies).  With the right modifications, this can be done.  But, the quantization is much smoother if we go back and revise the classical Dirac field theory that serves as the starting point for quantization---putting positrons into the theory and removing negative energies.

The revised picture of quantization put forward here modifies the starting point but leaves the endpoint essentially unchanged.  The goal is not to make any alterations to quantum field theory, but to improve our understanding of the foundations of the subject and to find a better way of presenting this material in textbooks and courses.

In section \ref{CDFT}, I give a quick introduction to classical Dirac field theory and then, in section \ref{howwasquantized}, I present a standard textbook-style approach to quantizing the Dirac field---where a number of modifications need to be made during quantization to ensure that the final theory includes positrons and excludes negative energies.  In the next section, I explain how we can revise the expressions for energy and charge in classical Dirac field theory (without altering the Dirac equation) so that the field always has positive energy, but can have either positive or negative charge.  In the following section, I show that the modifications made during quantization in section \ref{howwasquantized} are unnecessary if we start with the revised classical Dirac field theory of section \ref{revisingCDFT}.  The process of quantization becomes simple and straightforward.  In section \ref{comparisontoEM}, I compare the revised version of classical Dirac field theory to classical electromagnetism and show that it is in closer alignment than the original version of classical Dirac field theory.  I do not mention Dirac's idea that space is filled with an infinite sea of negative energy electrons until section \ref{diracsea}, where I explain how the revised version of classical Dirac field theory removes any temptation to think in such terms.  In this paper, we will view quantum field theory as built from a classical theory of fields, not particles.  In section \ref{minsize}, I briefly explore the implications of this picture (where the energy and charge of the electron are spread out) for understanding the self-interaction and spin of a classical electron.  We will see that the revised version of classical Dirac field theory from section \ref{revisingCDFT} yields a minimum size for the classical electron, large enough that we can understand the electron's angular momentum and magnetic moment as generated by true rotation (without needing any part of the electron to move faster than the speed of light).  In the main text of this article, I treat the classical Dirac field as complex-valued.  In appendix \ref{GrassmannCDFT}, I discuss the idea of a classical theory where the Dirac field is Grassmann-valued, explaining that such a theory could be used as a stepping stone between a classical theory where the Dirac field is complex-valued and a quantum theory of the Dirac field.

There are two very different approaches to quantum field theory: the particle approach and the field approach.  Focusing on the Dirac field, the field approach starts from classical Dirac field theory and then moves via field quantization\footnote{Although some authors speak of ``second quantization,'' this would not be an apt description of what is happening in either of the two approaches described above (neither one involves quantizing a theory that is already quantum).} to a quantum theory of the Dirac field.  The quantum state of the Dirac field is given by a wave functional\footnote{For an introduction to wave functionals, see \citet{jackiw1988, jackiw1990, hatfield}; \citet[ch.\ 11]{bohmhiley}; \citet[sec.\ 12.4]{holland}.  Wave functionals have been discussed by philosophers of physics in \citet{huggett2000, wallace2001, wallace2006, wallace2017, baker2009}; \citet[sec.\ 4.3.1]{myrvold2015}.  Wave functionals are called ``functionals'' because they are functions of functions (as the classical field configuration is itself a function).} that assigns amplitudes to possible configurations of the classical Dirac field.  The wave functional describes the quantum Dirac field as being in a superposition of different classical states.  Studying classical Dirac field theory helps us to better understand the elements of such superpositions.  According to the alternative particle approach,\footnote{The particle approach appears in, e.g., \citet[ch.\ 6--8]{schweberQFT}; \citet[sec.\ 13.2]{bjorkendrellfields}; \citet{thaller1992}; \citet[ch.\ 3]{teller}.} one starts instead from Dirac's single particle relativistic quantum theory of the electron and then moves to a quantum field theory by extending this single particle theory to multiple particles---permitting superpositions of different numbers of particles.\footnote{Adopting a Dirac sea picture, it may be possible to keep the number of particles fixed (see, e.g., \citealp{deckert}).}  In this approach, the quantum state is given by a wave function that assigns amplitudes to possible spatial arrangements of different numbers of point particles.   Both the particle and field approaches are pursued in the literature and there is much to be said about why one might prefer either approach over the other, or why one might prefer a different approach entirely.  In this paper I focus on understanding and developing the field approach.  But, it is not the only option.

\section{Classical Dirac Field Theory}\label{CDFT}

The process of field quantization starts from a classical theory of the Dirac field.  The classical Dirac field is sometimes treated as complex-valued and sometimes as Grassmann-valued.  In the main text of this article, I adopt the simpler picture of the classical Dirac field as complex-valued.  In appendix \ref{GrassmannCDFT}, I discuss the role that a Grassmann-valued classical Dirac field might play in the path from classical to quantum field theory.

Classical Dirac field theory describes the dynamics of the Dirac field, a field that carries both energy and charge.  The classical Dirac field $\psi_i(x)$ assigns four complex values (indexed by $i$) to each point in spacetime, $x$.  In what follows, the $i$ index on $\psi_i(x)$ will often be omitted and left implicit.  In the absence of interactions, the dynamics of the Dirac field is given by the free Dirac equation,
\begin{equation}
i\hbar \frac{\partial \psi}{\partial t}=\left(-i\hbar c \, \gamma^0 \vec{\gamma}\cdot\vec{\nabla}+ mc^2 \gamma^0 \right)\psi
\ .
\label{thediracequation}
\end{equation}
We can expand an arbitrary solution of \eqref{thediracequation} in terms of plane waves as
\begin{equation}
\psi (x) =\underbrace{\frac{1}{(2\pi\hbar)^{3/2}}\int{ \frac{d^3 p}{\sqrt{2 \mathcal{E}_p}} \sum_{s=1}^2 \left(b^s (p) \: u^s (p) \, e^{-\frac{i}{\hbar} p \cdot x}\right)}}_{\mbox{$\psi_+(x)$}}
+\underbrace{\frac{1}{(2\pi\hbar)^{3/2}}\int{ \frac{d^3 p}{\sqrt{2 \mathcal{E}_p}} \sum_{s=1}^2 \left(c^s (p) \: v^s (p) \,  e^{\frac{i}{\hbar} p \cdot x}\right)}}_{\mbox{$\psi_-(x)$}}
\ ,
\label{planewaveexpansion}
\end{equation}
where $\mathcal{E}_p = p^0 c = \sqrt{m^2 c^4 + |\vec{p}\,|^2 c^2}$.\footnote{In \eqref{planewaveexpansion} and throughout we use a ($+$ $-$ $-$ $-$) signature, as in \citet{bjorkendrell, bjorkendrellfields}; \citet{peskinschroeder}.}  In \eqref{planewaveexpansion}, the Dirac field $\psi$ is written as the sum of a positive frequency part $\psi_+$ and a negative frequency part $\psi_-$.  The positive and negative frequency parts are each expressed as a sum over polarizations $s$ and integral over 3-momenta $\vec{p}$, where each basis spinor with a particular polarization and momentum---$u^s (p)$ or $v^s (p)$---is assigned a complex amplitude---$b^s (p)$ or $c^s (p)$.  We can see that $\psi_+$ is associated with positive energy and $\psi_-$ with negative energy by writing the total energy of the Dirac field as\footnote{This total energy \eqref{totalenergy} may be viewed as the result of integrating either the energy density $i \hbar \, \psi^\dagger\frac{\partial \psi}{\partial t}$ (which appears in the canonical energy-momentum tensor) or the energy density $\frac{i \hbar}{2}\left(\psi^\dagger\frac{\partial \psi}{\partial t}-\frac{\partial \psi^\dagger}{\partial t}\psi\right)$ (which appears in the symmetrized energy-momentum tensor).  See \citet[pg.\ 419]{heitler}; \citet[pg.\ 219]{schweberQFT}.\label{twodensities}}
\begin{align}
\mathcal{E}&=\int{d^3 x \ i \hbar \, \psi^\dagger\frac{\partial \psi}{\partial t}}
\nonumber
\\
&=i \hbar \int{d^3 x \left( \psi_+^\dagger\frac{\partial \psi_+}{\partial t} + \psi_-^\dagger\frac{\partial \psi_-}{\partial t}\right)}
\nonumber
\\
&=\int{d^3 p \sum_{s=1}^2  \left(\mathcal{E}_p\,b^{s\dagger}(p)b^s(p)-\mathcal{E}_p\,c^{s\dagger}(p)c^{\,s}(p)\right)}
\ .
\label{totalenergy}
\end{align}
In \eqref{totalenergy} we've made use of the plane wave expansion in \eqref{planewaveexpansion} as well as the following properties of the basis spinors:
\begin{align}
u^{r\dagger} (p) u^s (p) &= v^{r\dagger} (p) v^s (p) = 2 \mathcal{E}_p \delta^{r s}
\nonumber
\\
u^{r\dagger} (p)v^s (p')&=v^{r \dagger} (p')u^s (p)=0\mbox{ \ \ if \ \ } \vec{p}\,'=-\vec{p}
\ .
\end{align}
The second property ensures that cross terms between positive and negative frequency modes in \eqref{totalenergy} drop out.

The (always negative) charge density of the classical Dirac field is
\begin{equation}
\rho^q=-e\, \psi^\dagger \psi
\ ,
\label{diracchargedensity}
\end{equation}
where the superscript $q$ indicates that this is a density of charge and $-e$ is the charge of the electron.  The current density describing the flow of this charge is
\begin{equation}
\vec{J}=-e c\, \psi^{\dagger} \gamma^0 \vec{\gamma} \psi
\label{chargecurrentcomplex}
\ .
\end{equation}

\section{How the Dirac Field has been Quantized}\label{howwasquantized}

With classical Dirac field theory on the table, we can now review a standard way\footnote{\citet{saunders1991} presents this standard quantization and finds similar shortcomings.  He advocates an alternative method of quantization (due to Segal) where complex numbers act differently on positive and negative frequency modes (see also \citealp{saunders1992, wallace2009}).} of quantizing the Dirac field (along the lines of \citealp[sec.\ 3.5]{peskinschroeder}; \citealp[ch.\ 5]{tong}).\footnote{My notation differs from theirs in factors of $2\pi$ because I write the plane wave expansion differently (and thus define the coefficients $b^s (p)$ and $c^s (p)$ differently).  With this notation, factors of $2\pi$ drop out in expressions for the energy and charge (as in, e.g., \citealp[ch.\ 8]{schweberQFT}; \citealp[sec.\ 13.4]{bjorkendrellfields}).  My notation also differs in that I include factors of $\hbar$ and $c$.}  

First, let us introduce the Heisenberg picture field operators $\widehat{\psi} (x)$, $\widehat{\psi}_+(x)$, and $\widehat{\psi}_-(x)$ by replacing the complex amplitudes in \eqref{planewaveexpansion} with annihilation operators,
\begin{equation}
\widehat{\psi} (x) = \underbrace{\frac{1}{(2\pi\hbar)^{3/2}}\int{ \frac{d^3 p}{\sqrt{2 \mathcal{E}_p}} \sum_{s=1}^2 \left(\widehat{b}^s (p) \: u^s (p) \, e^{-\frac{i}{\hbar} p \cdot x}\right)}}_{\mbox{$\widehat{\psi}_+(x)$}}
+\underbrace{\frac{1}{(2\pi\hbar)^{3/2}}\int{ \frac{d^3 p}{\sqrt{2 \mathcal{E}_p}} \sum_{s=1}^2 \left(\widehat{c}^{\,s} (p) \: v^s (p) \,  e^{\frac{i}{\hbar} p \cdot x}\right)}}_{\mbox{$\widehat{\psi}_-(x)$}}
\ .
\label{operatorplanewaveexpansion}
\end{equation}
Later, we will see that the coefficients $c^s (p)$ should actually have been replaced by creation operators for positrons.  But, for the moment we will put aside this bit of foreknowledge and continue with the expression above because (as we will see shortly) it fits naturally in a quantization of the classical field theory we started with (and also because it is instructive to see where this expression leads you astray).\footnote{Some authors include \eqref{operatorplanewaveexpansion} for pedagogical purposes (such as \citealp[sec.\ 8a]{schweberQFT}; \citealp[sec.\ 3.5]{peskinschroeder}; \citealp[sec.\ 5.3]{greiner1996}; \citealp[ch.\ 5]{tong}).  Others start from beginning with the positron creation and annihilation operators in \eqref{operatorswap} (such as \citealp[pg.\ 70]{hatfield}; \citealp[pg.\ 138]{ryder}; \citealp[pg.\ 211]{schwartz}).}

Replacing the field values in \eqref{totalenergy} with field operators, we arrive at a Hamiltonian for our quantum field theory,
\begin{align}
\widehat{H}&=\int{d^3 x \ i \hbar \, \widehat{\psi}^\dagger\frac{\partial \widehat{\psi}}{\partial t}}
\nonumber
\\
&=\int{d^3 p \sum_{s=1}^2  \left(\mathcal{E}_p\,\widehat{b}^{s\dagger}(p)\widehat{b}^s(p)-\mathcal{E}_p\,\widehat{c}^{\,s\dagger}(p)\widehat{c}^{\,s}(p)\right)}
\ .
\label{hamiltonian}
\end{align}
In \eqref{hamiltonian}, we see that the total energy is found by summing a number operator $\widehat{b}^{s\dagger}(p)\widehat{b}^s(p)$ times the energy $\mathcal{E}_p$ and a number operator $\widehat{c}^{\,s\dagger}(p)\widehat{c}^{\,s}(p)$ times the negative energy $-\mathcal{E}_p$.  In the classical field theory we started with, $\psi_+$ was associated with positive energy and negative charge, whereas $\psi_-$ was associated with negative energy and negative charge.  So, you might naturally think of $\widehat{b}^{s\dagger}(p)$ as the creation operator for a negatively charged electron in polarization state $s$ with momentum $\vec{p}$ and positive energy $\mathcal{E}_p$, and $\widehat{c}^{\,s\dagger}(p)$ as the creation operator for a negatively charged electron in polarization state $s$ with momentum $\vec{p}$ and negative energy $-\mathcal{E}_p$.\footnote{\citet[sec.\ 8a]{schweberQFT}; \citet[sec.\ 13.4]{bjorkendrellfields}; \citet{hatfield} take this interpretation of $\widehat{c}^{\,s\dagger}(p)$ quite seriously and retain it even after introducing $\widehat{d}^{s\dagger}(p)=\widehat{c}^{\,s}(p)$ as the creation operator for a positron with positive energy.  This Dirac-sea-style approach will be discussed in section \ref{diracsea}.}

Although it is possible to develop a quantum field theory with such negative energy particles, the resulting theory has a serious problem.  Electrons could emit an unlimited amount of radiation by dropping to states of arbitrarily low energy.  The standard move at this point is to reinterpret what appeared to be a theory of just electrons as a theory of both negatively charged electrons and positively charged positrons.\footnote{Another way of responding to this problem is to posit that the negative energy states are generally filled, so that (by Pauli exclusion) positive energy electrons are forbidden from dropping into arbitrarily low energy states (see section \ref{diracsea}).}  We can interpret $\widehat{c}^{\,s}(p)$ as a creation operator for positive energy particles with positive charge (positrons), revising our understanding of the vacuum so that it is  $\widehat{c}^{\,s\dagger}(p)$, not $\widehat{c}^{\,s}(p)$, that returns zero when acting on the vacuum.  To avoid confusion, let us introduce a new notation for these operators that allows us to retain the convention of writing daggers on creation operators,
\begin{equation}
\widehat{d}^{s\dagger}(p)=\widehat{c}^{\,s}(p)
\ .
\label{operatorswap}
\end{equation}
In this new notation, $\widehat{c}^{\,s\dagger}(p)\widehat{c}^{\,s}(p)$ becomes $\widehat{d}^{s}(p)\widehat{d}^{s\dagger}(p)$, which is clearly not a number operator.  To reorder these operators, we must make use of the anticommutation relations for the electron and positron creation and annihilation operators:
\begin{align}
\left\{\widehat{b}^{r}(p),\widehat{b}^{s \dagger}(q)\right\}&=\left\{\widehat{d}^{r}(p),\widehat{d}^{s \dagger}(q)\right\}= \delta^{rs}\delta^3(\vec{p}-\vec{q}\,)
\nonumber
\\
\left\{\widehat{b}^{r}(p),\widehat{b}^{s}(q)\right\}&=\left\{\widehat{d}^{r}(p),\widehat{d}^{s}(q)\right\}=...=\left\{\widehat{b}^{r}(p),\widehat{d}^{s \dagger}(q)\right\}=0
\ .
\label{anticommutation}
\end{align}
As a consequence,\footnote{See \citet[pg.\ 140]{ryder}.} the field operators obey the equal-time anticommutation relations
\begin{align}
\left\{\widehat{\psi}_i(\vec{x},t),\widehat{\psi}^{\dagger}_j(\vec{y},t)\right\}&= \delta_{ij}\delta^3 (\vec{x}-\vec{y})
\nonumber
\\
\left\{\widehat{\psi}_i(\vec{x},t),\widehat{\psi}_j(\vec{y},t)\right\}&=
\left\{\widehat{\psi}^{\dagger}_i(\vec{x},t),\widehat{\psi}^{\dagger}_j(\vec{y},t)\right\}=0
\ ,
\label{fieldanticommutation}
\end{align}
where here $i$ and $j$ index the four components of the Dirac field operators.\footnote{From properties of the basis spinors and the anticommutation relations for the creation and annihilation operators, one can derive anticommutation relations for $\widehat{\psi}_+$ and $\widehat{\psi}_-$ \citep[sec.\ 8b]{schweberQFT}.  Note that the anticommutators do not always vanish at spacelike separation.\label{anticommutationfootnote}}

There are a number of motivations that can be given for positing such anticommutation relations.\footnote{In addition to the motivations given above, one might also appeal to considerations of causality (\citealp[ch.\ 3]{peskinschroeder}; \citealp[sec.\ 5.5]{weinbergQFT}).}  One clear virtue of these relations is that they automatically ensure Pauli exclusion: it is impossible to create two electrons or two positrons in the same state since the creation operators anticommute with themselves and thus yield zero when applied twice.  Another commonly cited virtue is that these relations allow us to avoid negative energies.  We are about to see why that is thought to be true, though in our later approach to quantizing the Dirac field (section \ref{howtoquantize}) we will arrive at positive energies more directly and will not need to appeal to the anticommutation relations.

Using \eqref{anticommutation}, we can rewrite the Hamiltonian as
\begin{equation}
\widehat{H}=\int{d^3 p \sum_{s=1}^2  \left(\mathcal{E}_p\,\widehat{b}^{s\dagger}(p)\widehat{b}^s(p)+\mathcal{E}_p\,\widehat{d}^{s\dagger}(p)\widehat{d}^{s}(p)-\mathcal{E}_p \delta^3(0)\right)}
\ .
\label{fixedhamiltonian}
\end{equation}
Comparing \eqref{fixedhamiltonian} to \eqref{hamiltonian}, we see that this Hamiltonian associates positive energy with both electrons and positrons.  However, there is also an infinite contribution to the energy arising from the delta function in \eqref{anticommutation}.  We can remove this infinite term by redefining the Hamiltonian operator as
\begin{align}
\widehat{H}&=\int{d^3 p \sum_{s=1}^2 \left(\mathcal{E}_p\,\widehat{b}^{s\dagger}(p)\widehat{b}^s(p)+\mathcal{E}_p\,\widehat{d}^{s\dagger}(p)\widehat{d}^{s}(p)\right)}
\nonumber
\\
&=i \hbar \int{d^3 x \ :\widehat{\psi}^\dagger\frac{\partial \widehat{\psi}}{\partial t}:}
\nonumber
\\
&=i \hbar\int{d^3 x  \sum_{i=1}^4\left(\widehat{\psi}^{\dagger}_{+i}\frac{\partial \widehat{\psi}_{+i}}{\partial t}-\frac{\partial \widehat{\psi}_{-i}}{\partial t}\widehat{\psi}^{\dagger}_{-i} \right)}
\ ,
\label{fullyfixedhamiltonian}
\end{align}
where the double dots in the second line indicate normal-ordering (moving all annihilation operators to the right of all creation operators and inserting a minus sign whenever a fermion annihilation operator is moved past a creation operator).  In the third line, we enact this normal-ordering by writing the Hamiltonian in terms of the $\widehat{\psi}_{+}$ and $\widehat{\psi}_{-}$ operators in \eqref{operatorplanewaveexpansion}.

We can introduce a total charge operator for the Dirac field by replacing the field values in \eqref{diracchargedensity} with field operators and integrating over all of space,
\begin{align}
\widehat{Q}&= \int{d^3 x -e\,\widehat{\psi}^\dagger \widehat{\psi}}
\nonumber
\\
&=\int{d^3 p \sum_{s=1}^2  \left(-e\,\widehat{b}^{s\dagger}(p)\widehat{b}^s(p)+e\,\widehat{d}^{s\dagger}(p)\widehat{d}^{s}(p)-e\, \delta^3(0)\right)}
\ .
\label{badtotalcharge}
\end{align}
Here electrons are associated with negative charge and positrons with positive charge.  Just as with the operator for total energy (the Hamiltonian), the operator for total charge contains an infinite negative contribution arising from the anticommutation relations in \eqref{anticommutation}.  We can redefine the charge operator by dropping this infinite contribution,\footnote{This redefined charge operator appears in \citet[pg.\ 228]{schweberQFT}; \citet[pg.\ 60]{bjorkendrellfields}; \citet[pg.\ 71]{hatfield}; \citet[pg.\ 139]{ryder}; \citet[sec.\ 5.3]{greiner1996}; \citet[pg.\ 52]{duncan}.}
\begin{align}
\widehat{Q}&=\int{d^3 p \sum_{s=1}^2  \left(-e\,\widehat{b}^{s\dagger}(p)\widehat{b}^s(p)+e\,\widehat{d}^{s\dagger}(p)\widehat{d}^{s}(p)\right)}
\nonumber
\\
&=  \int{d^3 x \ -e :\widehat{\psi}^\dagger\widehat{\psi}:}
\nonumber
\\
&=  \int{d^3 x  \sum_{i=1}^4\left(-e\, \widehat{\psi}^{\dagger}_{+i} \widehat{\psi}_{+i}+e\, \widehat{\psi}_{-i}\widehat{\psi}^{\dagger}_{-i}\right)}
\ .
\label{totalcharge}
\end{align}

In this section we've seen that the road from classical Dirac field theory to quantum Dirac field theory is not direct.  We had to make a number of modifications: (i) swapping creation and annihilation operators \eqref{operatorswap} (and redefining the vacuum); (ii) changing the Hamiltonian from \eqref{hamiltonian} to \eqref{fullyfixedhamiltonian}; (iii) changing the charge operator from \eqref{badtotalcharge} to \eqref{totalcharge}.  The reason we had to make all of these modifications is that we started with a classical field theory of electrons (including negative energy modes) and moved to a quantum field theory of electrons and positrons (without negative energies).  \citet[pg.\ 138]{ryder} has written that the ``negative energy difficulty'' in classical Dirac field theory ``is only removed on quantisation.''  But, why must we wait until we've quantized the theory to correct it?  We can take the lessons that we learned in quantizing the Dirac field and apply them back to the classical field theory we started with, treating it as a field theory of both electrons and positrons.  In the next section, I will propose such a corrected version of classical Dirac field theory and then, in the following section, show that it provides a superior starting point for quantization (as we will not need to make the three modifications listed above).  The route to quantum Dirac field theory is smoother if we start with a classical field theory of both electrons and positrons (correcting the classical field theory first and then quantizing, instead of quantizing and then correcting).

\section{Revising Classical Dirac Field Theory}\label{revisingCDFT}

In this section and the next, I would like to start from the top and tell a new story where we understand from the beginning that we desire a theory of both electrons and positrons.  We begin with a new classical theory of the Dirac field and end with essentially the same quantum theory of the Dirac field.  Although the quantum field theory we arrive at will look just the way it did before and the empirical predictions of the theory will be unchanged, this new and more direct route will revise our understanding of the classical field states that enter superpositions in our quantum field theory.

As before, the classical Dirac field $\psi$ obeys the Dirac equation \eqref{thediracequation} and can be written as a sum of positive and negative frequency parts: $\psi=\psi_+ + \psi_-$.  Looking ahead to the quantum field theory that we would like to arrive at upon quantization and the operators for energy and charge that appear within it, \eqref{fullyfixedhamiltonian} and \eqref{totalcharge}, we can posit a new energy for the Dirac field in our revised classical field theory of\footnote{Because the field values commute in classical Dirac field theory, the energy in \eqref{newenergy0} could be rewritten more concisely as,
\begin{equation}
\mathcal{E}=i \hbar \int{d^3 x \left( \psi_+^\dagger\frac{\partial \psi_+}{\partial t} - \psi_-^\dagger\frac{\partial \psi_-}{\partial t}\right)}
\ .
\end{equation}
I chose to write \eqref{newenergy0} the way that I did because the order of these field values will matter when we quantize the theory (section \ref{howtoquantize}) and also because the order matters in Grassmann-valued classical Dirac field theory (appendix \ref{GrassmannCDFT}).} 
\begin{equation}
\mathcal{E}=i \hbar \int{d^3 x  \sum_{i=1}^4\left(\psi^{\dagger}_{+i}\frac{\partial \psi_{+i}}{\partial t}-\frac{\partial \psi_{-i}}{\partial t}\psi^{\dagger}_{-i} \right)}
\label{newenergy0}
\ ,
\end{equation}
and a new charge density of
\begin{equation}
\rho^q=\sum_{i=1}^4\left(-e\,\psi^{\dagger}_{+i} \psi_{+i}+e\,\psi_{-i}\psi^{\dagger}_{-i} \right)
\label{revisedchargedensity0}
\ .
\end{equation}
We can simplify the above expressions by introducing a new notation that divides the Dirac field into separate electron and positron fields.  Let us identify $\psi_+$ as the electron field $\psi_e$ and $\psi_-$ as the conjugate transpose of the positron field $\psi_p$,
\begin{align}
\psi_{e \, i}(x)&=\psi_{+ i}(x)
\nonumber
\\
\psi_{p \, i}(x)&=\psi^{\dagger}_{- i}(x)
\ .
\label{electronandpositronfields}
\end{align}
The total field $\psi$ in plane wave expansion is thus
\begin{equation}
\psi_i (x) =\underbrace{\frac{1}{(2\pi\hbar)^{3/2}}\int{ \frac{d^3 p}{\sqrt{2 \mathcal{E}_p}} \sum_{s=1}^2 \left(b^{s} (p) \: u_i^s (p) \, e^{-\frac{i}{\hbar} p \cdot x}\right)}}_{\mbox{$\psi_{e \, i}(x)$}}
+\underbrace{\frac{1}{(2\pi\hbar)^{3/2}}\int{ \frac{d^3 p}{\sqrt{2 \mathcal{E}_p}} \sum_{s=1}^2 \left(d^{s \dagger} (p) \: v_i^s (p) \,  e^{\frac{i}{\hbar} p \cdot x}\right)}}_{\mbox{$\psi^{\dagger}_{p \, i}(x)$}}
\ ,
\label{newplanewaveexpansion}
\end{equation}
where $b^{s} (p)$ are complex coefficients for the electron modes and $d^{s} (p)$ are complex coefficients for the positron modes.  Note that the possible configurations of the electron and positron fields are constrained by the fact that each has a plane wave expansion that includes only half of the modes available to the full Dirac field.

Rewriting the new charge density in \eqref{revisedchargedensity0} using \eqref{electronandpositronfields},
\begin{equation}
\rho^q=\underbrace{-e\, \psi_e^{\dagger} \psi_e}_{\mbox{$\rho^q_e$}} +\underbrace{e\, \psi_p^{\dagger} \psi_p}_{\mbox{$\rho^q_p$}}
\ ,
\label{revisedchargedensity}
\end{equation}
we see that it is the sum of a negative charge density from the electron field and a positive charge density from the positron field.  When you start considering interactions, this revision of the charge density---from \eqref{diracchargedensity} to \eqref{revisedchargedensity}---will modify the interaction between the classical Dirac and electromagnetic fields.  However, such a modification is welcome as it needs to be made at some point en route to quantum electrodynamics (because we ultimately desire a theory of both positive and negative charges).

Continuing with this pattern of redefinition, the charge current of the Dirac field becomes
\begin{equation}
\vec{J}=\underbrace{-e c\, \psi_e^{\dagger} \gamma^0 \vec{\gamma} \psi_e}_{\mbox{$\vec{J}_e$}} + \underbrace{e c\, \psi_p^{\dagger} \gamma^0 \vec{\gamma} \psi_p}_{\mbox{$\vec{J}_p$}}
\ .
\label{revisedchargecurrent}
\end{equation}
The charge density is locally conserved under the flow of charge prescribed by this current.  In fact, (in the free theory under consideration here) the charge densities associated with the electron and positron fields \eqref{revisedchargedensity} will each be independently locally conserved under their respective flows \eqref{revisedchargecurrent} (as follows directly from the fact that the positive and negative frequency pieces of $\psi$ each obey the Dirac equation).  One can combine $\rho^q_e$ and $\vec{J}_e$ into an electron four-current and $\rho^q_p$ and $\vec{J}_p$ into a positron four-current, each of which will transform properly under Lorentz transformations (which again can be seen as a result of the fact that $\psi_+$ and $\psi_-$ each obey the Dirac equation).

In this paper, we are treating $\psi$ as a classical field.  But, elsewhere $\psi$ is treated as a quantum wave function.  If you were thinking of $\psi$ as a wave function and seeking a probability density and current, \eqref{diracchargedensity} and \eqref{chargecurrentcomplex} would quickly yield viable candidates (since you could divide through by $-e$ and have an always positive probability that transforms properly under Lorentz transformations).  The density in \eqref{revisedchargedensity}, on the other hand, cannot be made positive simply by changing the constant out front.  However, because we are thinking of $\psi$ as a classical field and looking for a charge density (not a probability density), \eqref{revisedchargedensity} is entirely unproblematic.  Still, its discordance with the view of $\psi$ as a quantum wave function may help explain why the density and current in \eqref{revisedchargedensity} and \eqref{revisedchargecurrent} have not been proposed before.

The revised total energy of the Dirac field in \eqref{newenergy0} can be rewritten, using \eqref{electronandpositronfields} and moving the time derivative from $\psi_p^{\dagger}$ to $\psi_p$ (which flips the sign of that term), as\footnote{As was the case for the classical Dirac field theory of section \ref{CDFT} (see footnote \ref{twodensities}), the total energy in \eqref{newenergy} can be seen as a result of integrating the energy density
\begin{equation}
i \hbar \left( \psi_e^{\dagger}\frac{\partial \psi_e}{\partial t} + \psi_p^{\dagger}\frac{\partial \psi_p}{\partial t}\right)
\end{equation}
or as a result of integrating the alternative energy density
\begin{equation}
\frac{i \hbar}{2} \left( \psi_e^{\dagger}\frac{\partial \psi_e}{\partial t} -\frac{\partial\psi_e^{\dagger} }{\partial t}\psi_e+ \psi_p^{\dagger}\frac{\partial \psi_p}{\partial t}-\frac{\partial \psi_p^{\dagger}}{\partial t} \psi_p\right)
\ .
\end{equation}
\label{newdensities}
}
\begin{align}
\mathcal{E}&=i \hbar \int{d^3 x \left( \psi_e^{\dagger}\frac{\partial \psi_e}{\partial t} + \psi_p^{\dagger}\frac{\partial \psi_p}{\partial t}\right)}
\nonumber
\\
&=\int{d^3 p \sum_{s=1}^2  \left(\mathcal{E}_p\,b^{s\dagger}(p)b^{s}(p)+\mathcal{E}_p\,d^{s\dagger}(p)d^{s}(p)\right)}
\ .
\label{newenergy}
\end{align}
From the second line, it is immediately clear that both the electron and positron fields make positive contributions to the energy and that the total energy is conserved (as there is no time dependence).  Upon quantization, this energy will lead directly to the correct Hamiltonian operator \eqref{fullyfixedhamiltonian} (without negative energies or the problematic infinite term).

\section{How the Dirac Field Should be Quantized}\label{howtoquantize}

Starting with the revised classical Dirac field theory presented in the previous section streamlines the process of quantizing the Dirac field.  Moving from classical to quantum field theory, we can write the Dirac field operator in plane wave expansion as
\begin{equation}
\widehat{\psi}_i (x) = \underbrace{\frac{1}{(2\pi\hbar)^{3/2}}\int{ \frac{d^3 p}{\sqrt{2 \mathcal{E}_p}} \sum_{s=1}^2 \left(\widehat{b}^s (p) \: u_i^s (p) \, e^{-\frac{i}{\hbar} p \cdot x}\right)}}_{\mbox{$\widehat{\psi}_{e\,i}(x)$}}
+\underbrace{\frac{1}{(2\pi\hbar)^{3/2}}\int{ \frac{d^3 p}{\sqrt{2 \mathcal{E}_p}} \sum_{s=1}^2 \left(\widehat{d}^{\,s\dagger} (p) \: v_i^s (p) \,  e^{\frac{i}{\hbar} p \cdot x}\right)}}_{\mbox{$\widehat{\psi}^\dagger_{p\,i}(x)$}}
\ .
\end{equation}
by simply putting hats on the coefficients $b^s (p)$ and $d^{s\dagger} (p)$ in \eqref{newplanewaveexpansion} and viewing them as electron annihilation and positron creation operators (respectively).  There is no need to redefine creation and annihilation operators.  As before, we posit anticommutation relations for the creation and annihilation operators \eqref{anticommutation} and for the field operators \eqref{fieldanticommutation}.\footnote{From these anticommutation relations, one can calculate the anticommutation relations for $\widehat{\psi}_{e}$ and $\widehat{\psi}_{p}$ (see footnote \ref{anticommutationfootnote}).}

You get the correct Hamiltonian \eqref{fullyfixedhamiltonian} immediately from making the energy in \eqref{newenergy} into an operator expression,
\begin{align}
\widehat{H}&=i \hbar \int{d^3 x \left(\widehat{\psi}_e^{\dagger}\frac{\partial \widehat{\psi}_e}{\partial t} + \widehat{\psi}_p^{\dagger}\frac{\partial \widehat{\psi}_p}{\partial t}\right)}
\nonumber
\\
&=\int{d^3 p \sum_{s=1}^2  \left(\mathcal{E}_p\,\widehat{b}^{\, s\dagger}(p)\widehat{b}^{s}(p)+\mathcal{E}_p\,\widehat{d}^{\, s\dagger}(p)\widehat{d}^{\,s}(p)\right)}
\ .
\end{align}
Similarly, you get the correct charge operator \eqref{totalcharge} directly by swapping field values for field operators in \eqref{revisedchargedensity} and integrating over all of space,
\begin{align}
\widehat{Q}&= \int{d^3 x \left(-e\, \widehat{\psi}_e^\dagger\widehat{\psi}_e  + e\, \widehat{\psi}_p^\dagger\widehat{\psi}_p\right)}
\nonumber
\\
&=\int{d^3 p \sum_{s=1}^2  \left(-e\,\widehat{b}^{s\dagger}(p)\widehat{b}^s(p)+e\,\widehat{d}^{s\dagger}(p)\widehat{d}^{s}(p)\right)}
\ .
\end{align}

At this point, we have reached the conclusion of the main part of the paper.  I have put forward a new and more direct route to quantum Dirac field theory.  Next, we embark on three sections that explore different ramifications of this new approach to quantization.  In section \ref{comparisontoEM}, I compare the revised classical Dirac field theory of section \ref{revisingCDFT} to our classical theory of the electromagnetic field and show that it is in closer alignment than the old classical Dirac field theory of section \ref{CDFT}.  Thus far, I have avoided any talk of the Dirac sea in presenting either the original or the revised method of quantization.  In section \ref{diracsea}, I discuss the Dirac sea.  In section \ref{minsize}, I use the revised classical Dirac field theory to better understand the self-interaction and spin of the electron.

\section{Comparison to the Electromagnetic Field}\label{comparisontoEM}

At first glance, the expression for the energy of the electromagnetic field,
\begin{equation}
\mathcal{E}=\int{d^3 x \ \frac{1}{8\pi}(E^2 + B^2)}
\ ,
\label{EMfieldenergy}
\end{equation}
looks very different from either of the two different energies that have been proposed for the classical Dirac field, \eqref{totalenergy} or \eqref{newenergy}.  However, as we'll see shortly, there is an alternative way of expressing the state of the electromagnetic field that makes the energy of the electromagnetic field look very similar to these.  In particular, the signs will match the revised energy of the Dirac field in \eqref{newenergy0} and \eqref{newenergy}, where we associate positive energy with both positive and negative frequency modes.  This close parallel between electromagnetism and the revised classical Dirac field theory of section \ref{revisingCDFT} provides further reason to prefer this new version of classical Dirac field theory to the old version in section \ref{CDFT}.

One simple way to make the electromagnetic field look more like the Dirac field is to combine the electric and magnetic fields into a single complex vector field: $\vec{E}+i\vec{B}$.  However, you can do better by Fourier transforming that complex vector field, dividing by the square root of the energy per photon ($\hbar k c$, where $k=|\vec{k}|$), and Fourier transforming back (as is explained in \citealp{good1957, emasqp}).  We can thus express the state of the electromagnetic field using a three-component complex vector field $\phi$,\footnote{Although one could write the field in \eqref{expandingphi} as $\vec{\phi}$, we will use a notation close to that for $\psi$ by leaving the vector hat off but not forgetting that $\phi$ has three components (just as we remember that $\psi$ has four components even though the notation does nothing to remind us of this fact).} related to the electric and magnetic fields by
\begin{equation}
\phi_i(\vec{x},t)=\frac{1}{\sqrt{8\pi}}\frac{1}{(2\pi)^3}\int{ d^3 k  \frac{e^{i \vec{k}\cdot\vec{x}}}{\sqrt{\hbar k c}} \int{d^3 y\left(E_i(\vec{y},t)+iB_i(\vec{y},t)\right)e^{-i \vec{k}\cdot\vec{y}}} }
\label{expandingphi}
\ .
\end{equation}
From Maxwell's equations, one can derive an equation for the (free) time evolution of $\phi$ that is very similar to the Dirac equation \eqref{thediracequation},\footnote{Using the Levi-Civita symbol, the $\vec{s}$ matrices can be defined by the equation $(s_i)_{jk}=-i \epsilon_{ijk}$.}
\begin{equation}
i\hbar\frac{\partial \phi}{\partial t}  = -i \hbar c \: \vec{s}\cdot \vec{\nabla}\: \phi
\ .
\end{equation}
As with $\psi$, we can write $\phi$ as a sum of positive and negative frequency parts: $\phi=\phi_++\phi_-$.  \citet[eq.\ 31]{good1957} has shown that the energy of the electromagnetic field in \eqref{EMfieldenergy} can be written in terms of $\phi_+$ and $\phi_-$ as
\begin{align}
\mathcal{E}&=i \hbar \int{d^3 x \left(\phi^{\dagger}_+ \frac{\partial\phi_+}{\partial t} - \phi^{\dagger}_- \frac{\partial \phi_-}{\partial t} \right)}
\nonumber
\\
&=i \hbar \int{d^3 x \sum_{i=1}^3\left(\phi^{\dagger}_{+i} \frac{\partial\phi_{+i}}{\partial t} -  \frac{\partial \phi_{-i}}{\partial t} \phi^{\dagger}_{-i}\right)}
\ .
\label{phienergy}
\end{align}
As promised, this way of writing the energy of the electromagnetic field closely matches \eqref{newenergy0}, where there is a negative sign in front of the negative frequency modes ensuring that the energy is always positive.\footnote{The expression for the total energy in \eqref{phienergy} suggests a new energy density for the electromagnetic field, different from the standard $\frac{1}{8\pi}(E^2 + B^2)$:
\begin{equation}
i \hbar \left(\phi^{\dagger}_+ \frac{\partial\phi_+}{\partial t} - \phi^{\dagger}_- \frac{\partial \phi_-}{\partial t} \right)
\ .
\end{equation}
As was the case for the Dirac field (footnotes \ref{twodensities} and \ref{newdensities}), there is another possible energy density for the electromagnetic field,
\begin{equation}
\frac{i \hbar}{2} \left(\phi^{\dagger}_+ \frac{\partial\phi_+}{\partial t} -\frac{\partial\phi^{\dagger}_+}{\partial t} \phi_+ - \phi^{\dagger}_- \frac{\partial \phi_-}{\partial t} + \frac{\partial\phi^{\dagger}_-}{\partial t} \phi_-\right)
\ ,
\label{newEMenergydensity}
\end{equation}
that will yield the same total energy upon integration.  The energy density in \eqref{newEMenergydensity} is locally conserved (in the absence of charges) with respect to the energy flux density
\begin{equation}
\frac{i \hbar c^2}{2}\left(-\phi^{\dagger}_{+} \del \phi_{+} + (\del \phi^{\dagger}_{+}) \phi_{+}  + \phi^{\dagger}_{-} \del \phi_{-} - (\del \phi^{\dagger}_{-})\phi_{-} \right)
\label{newEMenergyfluxdensity}
\ ,
\end{equation}
to which one may add a curl term, as in \citet[eq.\ 13]{ohanian}.  The existence of the densities in \eqref{newEMenergydensity} and \eqref{newEMenergyfluxdensity} provides yet another illustration of the well-known fact that there are multiple ways to assign an energy density and energy flux density to the electromagnetic field while retaining local conservation of energy (\citealp[sec.\ 31-33]{landaulifshitzfields}; \citealp[sec.\ 6.7 and 12.10]{jackson}; \citealp[ch.\ 5]{lange}).}

Although it would be unorthodox to divide photons into particles and anti-particles, one could define particle and antiparticle fields as in \eqref{electronandpositronfields},
\begin{align}
\phi_{\gamma \, i}(x)&=\phi_{+ i}(x)
\nonumber
\\
\phi_{\bar{\gamma} \, i}(x)&=\phi^{\dagger}_{- i}(x)
\ ,
\end{align}
and rewrite the energy in \eqref{phienergy} as
\begin{equation}
\mathcal{E}=i \hbar \int{d^3 x \left( \phi_{\gamma}^{\dagger}\frac{\partial \phi_{\gamma}}{\partial t} + \phi_{\bar{\gamma}}^{\dagger}\frac{\partial \phi_{\bar{\gamma}}}{\partial t}\right)}
\ ,
\end{equation}
in analogy with \eqref{newenergy}.

\section{Draining the Dirac Sea}\label{diracsea}

In contemporary presentations of the quantization of the Dirac field, authors often seek to avoid any mention of Dirac's idea that space is filled with an infinite sea of negative energy electrons.  The Dirac sea has been described derisively as a ``potentially confusing metaphor" \citep[pg.\ 113]{zee2010}, ``an example of physicists not taking the trouble to rewrite their history'' \citep[pg.\ 120]{weinberg1985}, ``an extremely persistent distraction'' \citep[pg.\ 39]{duncan}, ``a red herring of sinister vitality'' \citep[pg.\ 34]{duncan}, and even ``total nonsense'' \citep[pg.\ 142]{schwartz}.  Being aware of this aversion to the Dirac sea, I have presented both the original and the revised methods of quantization (in sections \ref{howwasquantized} and \ref{howtoquantize}) without mentioning the Dirac sea.  However, there are other authors who put significant emphasis on the role of the Dirac sea in the quantization of the Dirac field (including \citealp[sec.\ 12]{heitler}; \citealp[sec.\ 8a]{schweberQFT}; \citealp[sec.\ 13.4]{bjorkendrellfields}; \citealp{hatfield}; \citealp[sec.\ 5.3]{greiner1996}).  As I see it, part of the confusion surrounding the Dirac sea comes from the fact that we start with a classical theory of only electrons (section \ref{CDFT}) and after quantization somehow want to arrive at a quantum field theory of electrons and positrons.  We can resolve this confusion by starting with a classical field theory of electrons and positrons (section \ref{revisingCDFT}).  In this section, I will first explain how one might see the Dirac sea as playing a role in the method of quantization from section \ref{howwasquantized} and then show that the revised method of quantization from section \ref{howtoquantize} removes any temptation to think in terms of a Dirac sea.

According to the Dirac sea picture, positrons are not fundamental.  At a deeper level, there are only electrons.  An absence of an electron in the negative energy Dirac sea (called a ``hole'') will act like a positively charged particle with positive energy and the same mass as the electron.  It will act like a positron.  Applying the idea that positrons are holes in the Dirac sea to the quantization of the Dirac field in section \ref{howwasquantized}, we can view the positron creation operator, $\widehat{d}^{s\dagger}(p)=\widehat{c}^{\,s}(p)$, as an operator that annihilates negative energy electrons (creating a hole in the sea) and the positron annihilation operator, $\widehat{d}^{s}(p)=\widehat{c}^{\,s\dagger}(p)$, as an operator that creates negative energy electrons (filling a hole in the sea).  The vacuum\footnote{\citet[pg.\ 70]{hatfield} calls this state the ``physical vacuum'' to distinguish it from the ``bare, truly empty, vacuum'' that contains no positive energy electrons and no negative energy electrons.  You might expect that we could use the same formalism to describe the physics of a few positive and negative energy electrons in an otherwise truly empty vacuum (recognizing that this imagined scenario is far from reality).  But, there is a problem.  If we assume that the anticommutation relations for the field operators are as in \eqref{fieldanticommutation}---derived in \eqref{derivingfieldcommutation}---then the creation and annihilation operators for negative energy electrons will obey the anticommutation relations $\left\{\widehat{c}^{\,r}(p),\widehat{c}^{\,s \dagger}(q)\right\}= - \delta^{rs}\delta^3(\vec{p}-\vec{q}\,)$.  A consequence of this is that single-negative-energy-electron states---formed by acting on the bare vacuum with a negative energy electron creation operator $\widehat{c}^{\,s \dagger}(q)$---will have negative norm.  \citet[sec.\ 5.1]{tong} mentions this kind of problem at the beginning of his quantization of the Dirac field.} is eigenstate of both $\widehat{b}^{s}(p)$ and $\widehat{c}^{\,s\dagger}(p)$ with eigenvalue zero.  Thus, it is a state in which all of the negative energy electron states are filled and all of the positive energy states are empty.  This picture yields a physical explanation for the infinities that appear in equations \eqref{fixedhamiltonian} and \eqref{badtotalcharge}.  The vacuum state will have infinite negative energy and infinite negative charge because there are infinitely many electrons with negative energy and negative charge in the Dirac sea.  We can understand our choice to remove these infinities by hand in section \ref{howwasquantized} as a choice to ignore this infinite background and focus on deviations from the vacuum state.

The perspective that I have just outlined has some attraction as a way of understanding the method of quantization in section \ref{howwasquantized} because it allows us to view the theory that we initially arrived at upon quantization as correct (even though it includes negative energy states and infinite terms in the energy and charge operators).  The modifications that were summarized at the end of section \ref{howwasquantized} can then be seen as mere changes in notation, not real alterations of the theory.\footnote{The infinities of \eqref{fixedhamiltonian} and \eqref{badtotalcharge} would be seen as physically real, but not worth carrying along with us as we calculate energies and charges in the theory.}  In the method of quantization presented in section \ref{howtoquantize}, we never encounter such infinite terms or negative energies.  By starting from a classical field theory of positive energy electrons and positrons, we bypass these oddities entirely.  As these features never arise, there is no reason to posit the existence of a sea of negative energy electrons to make sense of them.

In explaining their distaste for the Dirac sea, many authors have noted that the idea cannot be extended to bosons.\footnote{See, for example, \citet[pg.\ 119--120]{weinberg1985}; \citet[sec.\ 5.3]{tong}; \citet[pg.\ 142]{schwartz}.}  Although it may be possible to interpret the antiparticle of a particular fermion as a hole in the negative energy sea of such fermions, one cannot interpret the antiparticle of a particular boson as a hole in the negative energy sea of such bosons (because bosons do not obey the Pauli exclusion principle).  I agree with these authors that it would be best to explain the nature of antiparticles\footnote{For a variety of philosophical perspectives on the nature of antiparticles, see \citet{saunders1991, saunders1992, wallace2009, arntzenius2009, baker2010, deckert}.} and the removal of negative energies in the same way for bosons and fermions.  Negative energies are removed in the revised classical Dirac field theory by revising the expression for the energy so that negative frequency modes have positive energy, moving from \eqref{totalenergy} to \eqref{newenergy0}.  This revision also removes negative energies from the quantum field theory that you get after quantization (section \ref{howtoquantize}).  In the same way, the expression for the energy in \eqref{phienergy} ensures that there are no negative energies in our classical and quantum theories of the electromagnetic field.\footnote{\citet[pg.\ 1918]{good1957} mentions that the minus sign in \eqref{phienergy} cannot be explained by appeal to a negative energy sea because photons do not obey the Pauli exclusion principle.}

\section{The Classical Electron}\label{minsize}

In this paper, we are viewing our quantum field theory of positrons and electrons as built from a classical field theory, not a classical theory of point particles (see section \ref{introsec}).  The classical states that we are taking to enter superpositions in our quantum theory are classical field configurations, not arrangements of point particles.  From this perspective, what is a classical electron?  It is not a point particle, since the classical theory we are starting from is not a theory of point particles.  It is a field theory.

The classical Dirac field possesses both energy and charge.  Looking at the charge in the Dirac field provides a way of counting the number of particles that the field represents.  For the original classical Dirac field theory of section \ref{CDFT}, the number of electrons is the total charge---found by integrating \eqref{diracchargedensity} over space---divided by the charge of a single electron, $-e$ \citep[pg.\ 10]{takabayasi1957}.  For the revised classical Dirac field theory of section \ref{revisingCDFT}, the number of electrons can be found by dividing the charge of the electron field by $-e$,\footnote{This expression for the number of electrons is of the same form as the expression for the number of photons in \citet[pg.\ 1918]{good1957}; \citet[sec.\ 3]{emasqp}.}
\begin{equation}
\mbox{Number of Electrons}=\int{d^3 x \ \psi_e^{\dagger} \psi_e}
\ ,
\end{equation}
and the number of positrons can be found by dividing the charge of the positron field by $+e$.\footnote{Such calculations will not necessarily yield an integer number of electrons or positrons.}  A single classical electron would be described by a state in which the number of electrons is one (and the number of positrons is zero).  In such a state, the electron's energy and charge will be spread out, not located at a point.\footnote{\citet[pg.\ 54]{valentini1992} explains that the classical field states appearing in the quantum wave functional represent the electron as a ``field lump,'' not a point particle.  \citet{chuu2007} say that they ``regard the electron as a wavepacket in the positive energy spectrum of the Dirac equation.''  Along similar lines, \citet[pg.\ 77]{weinberg2018} explains the status of particles in quantum field theory for a general audience by writing: ``From the perspective of quantum field theory, as developed by Heisenberg, Pauli, Victor Weisskopf and others in the period 1926--1934, the basic ingredients of nature are not particles but fields.  Particles like the electron and photon are bundles of energy of the electron field and the electromagnetic field, respectively.''}

This picture of the classical electron as extended is attractive for a couple of reasons: (i) we avoid problems with point charges in classical electromagnetism, and (ii) we are able to think of the electron as truly rotating and not as a point particle that somehow possesses intrinsic angular momentum and magnetic moment.  Let's discuss (i) first.  In classical electromagnetism we often treat charged particles as point-size, but this leads to problems when calculating the force that a charged particle feels from the electromagnetic field that it itself produces (as this field is infinitely strong at the location of the particle).  One cannot simply ignore this self-force because it has been observed in experiment and must be included to achieve conservation of energy and momentum.  Detailed discussions of these problems with self-interaction can be found in many textbooks on classical electromagnetism.  Philosophers of physics are aware of these issues and have considered a number of possible responses \citep{lange, frisch2005, lazarovici2018, maudlin2018, hartensteinhubert}.  One way out is to think of the classical electron as extended.  This route was taken by Abraham, Lorentz, and Poincar\'{e} (see \citealp{pearle}).  But, their models have not been incorporated into contemporary physics.  The picture of the classical electron as extended that is being examined here is quite different.  If one takes a field approach to understanding quantum field theory, then quantum electrodynamics can be seen as already built upon such a classical picture.\footnote{Thinking of the electron as extended removes the infinities associated with self-interaction in the classical theory, but it prompts another question: What holds the electron together in the face of the electric repulsion that is threatening to tear it apart?  One could posit additional forces (see the Poincar\'{e} stresses in \citealp[ch.\ 16]{jackson}), but looking within our classical theories of the Dirac and electromagnetic fields there seems to be nothing holding the electron together.  Perhaps stability only comes when we move from the classical field theory to a quantum one.  This would be an interesting area for further research.}

The second attraction of this picture is that it allows us to understand the electron as truly spinning.  It is often claimed that one cannot regard the electron as actually rotating (or in a quantum superposition of states in which it is actually rotating) because the electron is too small.  If it were rotating, its edges would have to be moving faster than the speed of light in order for it to have the correct angular momentum and magnetic moment.\footnote{See \citet[pg.\ 47]{uhlenbeck}; \citet[pg.\ 35]{tomonaga}; \citet[problem 4.25]{griffithsQM}; \citet[pg.\ 127]{rohrlich}; \citet[sec.\ 2]{howelectronsspin}.}  But, how small is the electron?

In the classical Dirac field theory of section \ref{CDFT}, there is no limit to how tightly peaked the energy and charge densities might be when the total electron number is one.  The classical electron can be arbitrarily small.  However, that is not the case for the revised classical Dirac field theory of section \ref{revisingCDFT}.  In that theory, a classical field configuration for a single electron (and no positrons) must be composed entirely out of positive frequency modes of the Dirac field.  One cannot construct an arbitrarily tightly peaked state for the classical electron out of such modes.  There is a minimum size over which the energy and charge of the electron must be spread, on the order of the Compton radius $\frac{\hbar}{m c}$ (\citealp{newton1949}; \citealp[pg.\ 299]{heitler}; \citealp[pg.\ 39]{bjorkendrell}).  This is the minimum size that is needed in order to avoid superluminal rotation.  So, the classical electron that enters superpositions in quantum field theory is large enough that it could be spinning.  By examining the flow of energy and charge in the classical electron field, we can see that the electron is actually spinning \citep{ohanian, chuu2007, howelectronsspin}.  Further, we can explain why the electron's gyromagnetic ratio is off by a factor of two from the simplest classical estimate: the charge in the Dirac field rotates twice as fast as the energy \citep{howelectronsspin}.

Of course, actual electrons are not accurately described by classical Dirac field theory.  In quantum electrodynamics, the physical state of a lone electron would be given by a wave functional that assigns complex amplitudes to classical configurations of the electron, positron, and photon fields.  To better understand this full quantum description of the electron, it is helpful to first study the classical field configurations and their classical dynamics.  In this section, we have analyzed self-interaction and spin in the simple case of a pure classical electron field.  This analysis lays the groundwork for a deeper understanding of self-interaction and spin in quantum field theory where the classical electron, positron, and photon fields enter quantum superpositions.

\section{Conclusion}

The Dirac field is sometimes called ``the electron-positron field.''  But, in the ordinary approach to quantizing the Dirac field, positrons only enter after the classical Dirac field has been quantized.  In this paper, I have presented a new approach to quantizing the Dirac field in which we start from the beginning with a classical field theory of electrons and positrons.  In this theory, the classical Dirac field can be decomposed into separate electron and positron fields---where the former has negative charge and the latter has positive charge, but both have positive energy.  Starting with this revised version of classical Dirac field theory streamlines the quantization of the Dirac field, brings our physics of the electron and positron into closer parallel with our physics of the photon, and yields an improved understanding of electron spin.\\

\vspace*{12 pt}
\noindent
\textbf{Acknowledgments}
Thank you to Sean Carroll, Dustin Lazarovici, David Wallace, and the anonymous reviewers for helpful feedback and discussion.  Special thanks to Jacob Barandes for pointing out an important problem in an earlier draft.\\

\appendix
\section{Grassmann-Valued Classical Dirac Field Theory}\label{GrassmannCDFT}

In the main text of this article, I sketched a simple picture of the path from classical to quantum field theory that started from a classical theory of the Dirac field (where the Dirac field is complex-valued) and moved directly to a quantum theory of the Dirac field through a procedure of field quantization.  In a more detailed picture of the path from classical to quantum field theory, we may want to insert an intermediate step: moving first from a complex-valued to a Grassmann-valued classical theory of the Dirac field and then quantizing the Grassmann-valued theory to arrive at a quantum theory of the Dirac field.  In this appendix, I present Grassmann-valued classical Dirac field theory, explain its relation to complex-valued classical Dirac field theory, and discuss the role it might play in the quantization of the Dirac field.

To see a reason for introducing a Grassmann-valued classical Dirac field, let us examine the action of field operators on states in quantum field theory.  We can represent the quantum state of the Dirac field (in the Schr\"{o}dinger picture) by a time-dependent wave functional $\Psi[\psi,t]$ that assigns at every time an amplitude to each possible configuration of the classical Dirac field and evolves via a Schr\"{o}dinger equation,
\begin{equation}
i \hbar \frac{\partial}{\partial t}\Psi[\psi,t]=\widehat{H}\Psi[\psi,t]
\ .
\end{equation}
The field operators $\widehat{\psi}_i(\vec{x})$ and $\widehat{\psi}_i^{\dagger}(\vec{x})$ act on the wave functional as follows:\footnote{See \citet[pg.\ 217]{hatfield}.}
\begin{align}
\widehat{\psi}_i(\vec{x})\Psi[\psi,t]&=\psi_i(\vec{x})\Psi[\psi,t]
\nonumber
\\
\widehat{\psi}_i^{\dagger}(\vec{x})\Psi[\psi,t]&=\frac{\delta}{\delta \psi_i(\vec{x})}\Psi[\psi,t]
\label{fieldoperatorsonWFs}
\ .
\end{align}
Here $\widehat{\psi}_i(\vec{x})$ multiplies the wave functional by the value of the $i$-th component of the classical Dirac field at $\vec{x}$ and $\widehat{\psi}_i^{\dagger}(\vec{x})$ takes the functional derivative of the wave functional with respect to $\psi_i(\vec{x})$.  If the classical field values were complex, then the anticommutation relations in \eqref{fieldanticommutation} would not be consistent with \eqref{fieldoperatorsonWFs} (the field operators $\widehat{\psi}_i(\vec{x})$ and $\widehat{\psi}_j(\vec{y})$ would commute because the field values would commute).  Instead, we can take the classical field values that appear in \eqref{fieldoperatorsonWFs} to be anticommuting Grassmann numbers.\footnote{It is standard practice to treat the classical Dirac field as Grassmann-valued when calculating path integrals.  For an introduction to the mathematics of Grassmann numbers and their application in quantum field theory, see \citet{berezin}; \citet{hatfield}; \citet[ch.\ 4]{valentini1992}; \citet{valentini1996}; \citet[sec.\ 9.5]{peskinschroeder}; \citet[sec.\ 6.7]{ryder}; \citet[sec.\ 12.8]{greiner1996}; \citet[sec.\ 11.5]{zee2010}; \citet[sec.\ 10.3.2]{duncan}; \citet[sec.\ 14.6]{schwartz}.}  Then, the correct anticommutation relations \eqref{fieldanticommutation} follow from \eqref{fieldoperatorsonWFs}.\footnote{Here we calculate these anticommutation relations in the Schr\"{o}dinger picture, assuming the Dirac field is Grassmann-valued and making use of the fact that $\left\{\frac{\delta}{\delta \psi_j(\vec{y})},\psi_i(\vec{x})\right\}=\delta_{ij}\delta^3 (\vec{x}-\vec{y})$ (see \citealp[eq.\ 9.63]{hatfield}):
\begin{align}
\left\{\widehat{\psi}_i(\vec{x}),\widehat{\psi}^{\dagger}_j(\vec{y})\right\}\Psi[\psi,t]&=\left(\psi_i(\vec{x})\frac{\delta}{\delta \psi_j(\vec{y})}+\frac{\delta}{\delta \psi_j(\vec{y})}\psi_i(\vec{x})\right)\Psi[\psi,t]=\delta_{ij}\delta^3 (\vec{x}-\vec{y})\Psi[\psi,t]
\nonumber
\\
\left\{\widehat{\psi}_i(\vec{x}),\widehat{\psi}_j(\vec{y})\right\}\Psi[\psi,t]&=\left(\psi_i(\vec{x})\psi_j(\vec{y})+\psi_j(\vec{y})\psi_i(\vec{x})\right)\Psi[\psi,t]=0
\nonumber
\\
\left\{\widehat{\psi}^{\dagger}_i(\vec{x}),\widehat{\psi}^{\dagger}_j(\vec{y})\right\}\Psi[\psi,t]&=\left(\frac{\delta}{\delta \psi_i(\vec{x})}\frac{\delta}{\delta \psi_j(\vec{y})}+\frac{\delta}{\delta \psi_j(\vec{y})}\frac{\delta}{\delta \psi_i(\vec{x})}\right)\Psi[\psi,t]=0
\label{derivingfieldcommutation}
\ .
\end{align}
}  If we take the classical Dirac field to be Grassmann-valued, then it appears that we must also take the wave functional to be Grassmann-valued.

Recognizing that we want to be quantizing a Grassmann-valued classical Dirac field to arrive at our quantum theory of the Dirac field, we can view the route to quantum field theory as beginning with the relatively easy-to-understand complex-valued classical Dirac field theory, then moving to Grassmann-valued classical Dirac field theory, and finally continuing on to quantum field theory by a process of field quantization.  Because we have introduced two distinct complex-valued classical Dirac field theories in sections \ref{CDFT} and \ref{revisingCDFT}, there are two different potential starting points for this route.  Although the revised classical Dirac field theory of section \ref{revisingCDFT} is the better choice, we will begin with the standard complex-valued classical Dirac field theory of section \ref{CDFT} and consider the transition from this theory to a Grassmann-valued classical theory of the Dirac field.

Let us introduce an infinite collection of distinct Grassmann numbers, $\alpha_i(\vec{x})$, such that four Grassmann numbers (indexed by $i$) are associated with each point in space,\footnote{The basis fields are taken to be functions of the three-vector $\vec{x}$, not the four-vector $x$, so that the space of possible configurations for the Grassmann-valued Dirac field \eqref{complextoGrassmann} remains constant over time.  The wave functional assigns time-dependent complex amplitudes over a fixed space of possible configurations for the classical Dirac field.} $\vec{x}$.  Let us also introduce an infinite collection, $\alpha^*_i(\vec{x})$, of conjugates for these Grassmann numbers.\footnote{The $\alpha$ notation is adapted from \citet[pg.\ 62--67]{berezin}.}  The Grassmann numbers in these two infinite collections will serve as a basis for writing other Grassmann numbers.  (The numbers in these collections are the generators of our Grassmann algebra.)  The basis fields $\alpha_i(\vec{x})$ and $\alpha^*_i(\vec{x})$ obey the anticommutation relations:
\begin{align}
\{\alpha_i(\vec{x}),\alpha_j(\vec{y})\}&=0
\nonumber
\\
\{\alpha^*_i(\vec{x}),\alpha_j(\vec{y})\}&=0
\nonumber
\\
\{\alpha^*_i(\vec{x}),\alpha^*_j(\vec{y})\}&=0
\label{alphaanticommutation}
\end{align}

Using these basis fields, we can write the Grassmann-valued Dirac field $\psi^G(x)$ in terms of the complex-valued Dirac field $\psi^c(x)$ as
\begin{equation}
\psi^G_i(x)=\psi^c_i(x) \alpha_i(\vec{x})
\ .
\label{complextoGrassmann}
\end{equation}
Here and henceforth, I will use the superscripts $G$ and $c$ to distinguish the complex-valued and Grassmann-valued classical Dirac fields.  Via \eqref{complextoGrassmann}, we can go back and forth between the complex-valued and Grassmann-valued fields: multiply $\alpha_i(\vec{x})$ times $\psi_i^c(x)$ to get $\psi^G_i(x)$, or, find the complex factor multiplying $\alpha_i(\vec{x})$ in $\psi^G_i(x)$ to get $\psi_i^c(x)$.  There is a one-to-one mapping between the states in each theory.  Because of this one-to-one mapping, we can equivalently think of the wave functional as assigning complex amplitudes to possible configurations of either the complex-valued or Grassmann-valued classical Dirac field.\footnote{This means that whether or not we choose to insert the intermediate step of Grassmann-valued classical Dirac field theory between complex-valued Dirac field theory and quantum field theory, we can view the quantum state of the Dirac field as a superposition of states of the complex-valued classical Dirac field.  This is the view that was adopted in the main text and that played an important role in section \ref{minsize}.}  We can also use \eqref{complextoGrassmann} to see that if $\psi^c(x)$ obeys the Dirac equation, $\psi^G(x)$ will as well---provided we understand the spatial derivatives to act on $\psi^c(x)$ and not the basis fields $\alpha_i(\vec{x})$.  Note that the complex-valued Dirac field in \eqref{complextoGrassmann} can be written as a sum of positive and negative frequency parts, related to the positive and negative frequency parts of the Grassmann-valued Dirac field by $\psi^G_{+i}(x)=\psi_{+i}^c(x) \alpha_i(\vec{x})$ and $\psi^G_{-i}(x)=\psi_{-i}^c(x) \alpha_i(\vec{x})$.

We can introduce a charge density for the Grassmann-valued Dirac field by adding $G$ superscripts to \eqref{diracchargedensity},
\begin{align}
\rho^q(x)&=-e\, \psi^{G\dagger}(x) \psi^G(x)
\nonumber
\\
&= -e \sum_{i=1}^{4}\psi^{c\dagger}_i(x)\psi^c_i(x) \alpha^{*}_i(\vec{x})\alpha_i(\vec{x})   
\ .
\label{diracchargedensityGrassmann}
\end{align}
The factor of $-e$ out front is negative.  But, in Grassmann-valued classical Dirac field theory, the charge density \eqref{diracchargedensityGrassmann} is not real-valued and thus it is neither positive nor negative. It is Grassmann-valued.\footnote{It would be convenient if $\alpha_i^*(\vec{x})\alpha_i(\vec{x})$ were equal to 1, as the charge densities for the complex-valued and Grassmann-valued Dirac fields would both be real-valued and would be equal to one another.  However, it cannot be.  By \eqref{alphaanticommutation}, the square of $\alpha_i^*(\vec{x})\alpha_i(\vec{x})$ is zero (not 1):
\begin{equation}
\alpha_i^*(\vec{x})\alpha_i(\vec{x})\alpha_i^*(\vec{x})\alpha_i(\vec{x})=-\alpha_i^*(\vec{x})\alpha_i^*(\vec{x})\alpha_i(\vec{x})\alpha_i(\vec{x})=0
\ .
\end{equation}
\citet[pg.\ 6]{dewitt1992} and \citet[pg.\ 24]{rogers2007} postulate that a basis element $\beta$'s conjugate, $\beta^*$, is either equal to $\beta$ (DeWitt) or $i \beta$ (Rogers).  Either way, the product of a basis element's conjugate with itself is zero.  That consequence makes these choices unwise for our purposes here.  If $\alpha_i^*(\vec{x})\alpha_i(\vec{x})$ were equal to zero, then the energy and charge of the Dirac field would always be zero regardless of the state of the field---see \eqref{diracchargedensityGrassmann} and \eqref{classicalenergyGrassmann}.  I follow \citet[pg.\ 66--67]{berezin} in leaving the product of a basis element's conjugate with itself unspecified.  (This aligns with common practice in quantum field theory textbooks where only anticommutation relations are postulated.  See, e.g., \citealp[sec.\ 10.3.2]{duncan}; \citealp[sec.\ 14.6]{schwartz}.)}  We can similarly introduce a Grassmann-valued energy for the Dirac field by adding $G$ superscripts to \eqref{totalenergy},\footnote{If we reorder the negative frequency field values in the first line of \eqref{classicalenergyGrassmann}, we can rewrite this energy as
\begin{equation}
\mathcal{E}=i \hbar\int{d^3 x \sum_{i=1}^{4}  \left( \psi_{+i}^{G\dagger}\frac{\partial \psi^G_{+i}}{\partial t} - \frac{\partial \psi^G_{-i}}{\partial t}\psi_{-i}^{G\dagger}\right)}
\label{fullyfixedenergy}
\ ,
\end{equation}
which would naturally yield the correct Hamiltonian operator \eqref{fullyfixedhamiltonian} upon quantization.  However, it is unclear why that form should be preferred in this classical theory of the Dirac field.}
\begin{align}
\mathcal{E}&=i \hbar \int{d^3 x \left( \psi_+^{G\dagger}\frac{\partial \psi^G_+}{\partial t} + \psi_-^{G\dagger}\frac{\partial \psi^G_-}{\partial t}\right)}
\nonumber
\\
&=i \hbar \int{d^3 x \ \psi^{G\dagger}\frac{\partial \psi^G}{\partial t}}
\nonumber
\\
&=i \hbar \int{d^3 x \sum_{i=1}^{4} \left( \psi_i^{c \dagger} (x) \frac{\partial \psi_i^c(x)}{\partial t} \alpha^{*}_i(\vec{x})\alpha_i(\vec{x}) \right)}
\ .
\label{classicalenergyGrassmann}
\end{align}
It is strange (and perhaps unacceptable) to have a classical field with Grassmann-valued energy and charge.  One issue is that interactions between this field and others are problematic.  For the complex-valued Dirac field, it is easy to see how the field's real-valued charge could interact with the classical electromagnetic field via Maxwell's equations.  However, it is hard to see how we could have a coherent theory where the Grassmann-valued Dirac field's Grassmann-valued charge interacts with the classical electromagnetic field.

Some authors have questioned whether what I've called ``Grassmann-valued classical Dirac field theory'' deserves to be called a ``classical'' theory.  \citet[pg.\ 28]{bailinlove} point out that moving to a Grassmann-valued Dirac field ``evades the positivity problem [i.e., the problem of negative energies] by making the energy a Grassmann variable, rather than a real number, and consequently not something whose positivity, or lack of it, can be enquired about.''  Immediately after that, they conclude: ``Thus it is clear that in this approach spin fields are essentially non-classical.''  Along similar lines, \citet{casalbuoni1976} has suggested that we call a system ``pseudoclassical'' if it is ``described by ordinary c-number variables ... and by Grassmann variables'' (see also \citealp[ch.\ 11]{freund1986}).  I agree with these authors that Grassmann-valued classical Dirac field theory is very different from our classical theories of the electromagnetic and gravitational fields.  But, I have chosen to retain the word ``classical'' to indicate that, whatever this theory is, it is not a quantum theory.

The process of quantizing the above Grassmann-valued classical theory of the Dirac field would proceed in much the same way\footnote{One advantage of this new approach to quantization is that the anticommutation relations for the field operators (and for the creation and annihilation operators) could be derived from \eqref{fieldoperatorsonWFs} and would not have to be postulated.} as in section \ref{howwasquantized} and would run into the same problems.  We would do better to revise our Grassmann-valued classical Dirac field theory so that it matches our revised complex-valued classical Dirac field theory from section \ref{revisingCDFT}.  We can separate the Grassmann-valued classical Dirac field into separate Grassmann-valued electron and positron fields, related to the complex-valued fields by
\begin{align}
\psi^G_{e\,i}(x)&=\psi^c_{e\,i}(x) \alpha_i(\vec{x})
\nonumber
\\
\psi^G_{p\,i}(x)&=\psi^c_{p\,i}(x) \alpha_i(\vec{x})
\ .
\end{align}
In this revised Grassmann-valued classical Dirac field theory, the Grassmann-valued charge density would be given by \eqref{revisedchargedensity} and the Grassmann-valued energy by \eqref{newenergy} (with $G$ superscripts inserted).  Quantizing this revised theory would proceed smoothly, as in section \ref{howtoquantize}.

In section \ref{introsec}, I briefly explained two different approaches to quantum field theory: the particle approach and the field approach.  From the perspective of someone working to extend Bohmian mechanics to relativistic quantum field theory, the distinction between particle and the field approaches is especially important.  These approaches suggest different additions to the quantum state: particles or fields.  There has been disagreement among Bohmians over whether one should take a particle or field approach for fermions (like the electron and positron).  One problem that has been raised for the field approach is that the fields may have to be Grassmann-valued.  Some have seen this as as a serious issue and others have thought it unproblematic.  Compare \citet[pg.\ 374]{bohm1987}; \citet[pg.\ 156]{kaloyerou1996}; \citet[sec.\ 9.2]{struyve2010}; \citet[sec.\ 3.3]{struyve2011} to \citet[sec.\ 4.2]{valentini1992}; \citet{valentini1996}; \citet[pg.\ 519]{holland}.

The above analysis of Grassmann-valued classical Dirac field theory can be used to address some of the problems that have been raised by these authors.  \citet[pg.\ 374]{bohm1987} write: ``... if we regard the fermions as fields, they obey anticommutation relations which have no classical limit and which do not permit a picture of continuous field variables that we have used for bosonic systems.''  As was mentioned in the introduction, classical Dirac field theory is not the classical limit of quantum Dirac field theory.  Still, classical Dirac field theory has a role to play in arriving at quantum Dirac field theory.  It is possible for a classical field to obey anticommutation relations if the field is Grassmann-valued.  Looking at \eqref{complextoGrassmann}, the Grassmann-valued Dirac field inherits a kind of continuity from the fact that the complex-valued Dirac field varies continuously in space and time.  Responding to \citet{valentini1992, valentini1996}, \citet[sec.\ 9.2.2]{struyve2010} raises concerns for interpreting a quantity like $\Psi^*[\psi^G,t]\Psi[\psi^G,t]$ as a probability density over the space of possible field configurations.  Struyve writes that ``it is unclear what measure should be considered on the space of anti-commuting fields.''  Because of the one-to-one mapping between states of the complex-valued and Grassmann-valued Dirac fields in \eqref{complextoGrassmann}, we can use the same measure for the Grassmann-valued field as for the complex-valued field.  Struyve also objects that if the wave functional is Grassmann-valued, then the probability density will be Grassmann-valued as well.  That indeed seems to be a serious problem facing the use of wave functionals for Grassmann-valued fields.

\end{document}